# An ion species model for positive ion sources – part I description of the model

E Surrey[1] and A J T Holmes[2]

[1]EURATOM/CCFE Fusion Association, Culham Science Centre, Abingdon, Oxfordshire OX14 3DB, UK

[2]Marcham Scientific, Hungerford, Berkshire RG17 0LH, UK

*Abstract*. A one dimensional model of the magnetic multipole volume plasma source has been developed for use in intense ion/neutral atom beam injectors. The model uses plasma transport coefficients for particle and energy flow to create a detailed description of the plasma parameters along an axis parallel to that of the extracted beam. Primarily constructed for applications to neutral beam injection systems on fusion devices, the model concentrates on the hydrogenic isotopes but can be extended to any gas by substitution of the relevant masses, cross sections and rate coefficients. The model considers the flow of fast ionizing electrons that create the ratios of the three hydrogenic isotope ion species, $H^+$, $H_2^+$, $H_3^+$ (and similarly for deuterium and tritium) as they flow towards the beam extraction electrode, together with the production of negative hydrogenic ions through volume processes. The use of detailed energy balance in the discharge allows the determination of the fraction of the gas density that is in an atomic state and also the gas temperature as well as the electron temperatures and plasma potential. Comparisons are made between the results of the model and experimental measurements in deuterium from a number of different filament driven sources used on beam heating facilities.

**Keywords:** positive ion source, ion species, plasma modelling

**PACS:** 41.75.Cn, 52.50.Dg, 52.40.Kh

## 1. Introduction

Neutral beam injection is a major method of providing non-inductive current drive and heating to magnetically confined fusion plasma devices. The beams, generally of $D^0$ atoms, are formed from either positive or negative ion precursors, the choice being dictated by the beam energy required for plasma penetration. For positive ions, the neutral beam is created by charge exchange in a gas cell for which the neutralisation efficiency is less than 30% at beam energy above ~75keV/amu. For negative ions the neutral beam is created by electron stripping in a gas cell, for which the efficiency has a maximum value of ~58% for all energies above ~50keV/amu. The plasma generator that forms the precursor ions is of the magnetic multipole type and may contain a magnetic "filter" to create a low temperature plasma in the region from which the beam is extracted. This technique, which reduces the electron energy in the extraction region, has different consequences for positive and negative ion production. For positive ion sources it serves to reduce the direct ionisation of the neutral gas, limiting the $D_2^+$ production and promoting the dissociation of $D_3^+$ into $D^+$ and $D_2$ [1]. In the negative ion source the electron energy is reduced to an energy at which electron detachment collisions are no longer dominant [2] so that the production of $D^-$ is maximised. Both of these effects occur at an electron temperature of approximately 1eV. The negative ions are formed by either dissociative attachment collisions (also maximised at about 1eV temperature) or by emission from a caesiated surface.

Despite its history the role of this magnetic filter is not clearly understood and it is the intention of this paper to develop a model for this purpose. The magnetic filter itself is extremely simple; it is a sheet of magnetic field dividing the source chamber into two essentially field free regions - the one next to the beam extraction apertures referred to as the extraction region and the other at the back of the source chamber referred to as the driver region. This latter region also contains the discharge excitation system. This can be an RF drive antennae but for many discharges it is a DC hot wire filament. This paper will focus on the latter technique as most of the data presented here is from sources of this type.

The first use of a magnetic filter was Ehlers and Leung [3] and Holmes et al. [4, 5, 6] where it was used to enhance the production of $H^+$ ions relative to the molecular ions $H_2^+$ and $H_3^+$. The effect was attributed to the dissociation of the molecular ions by cold electron impact while the lack of fast electrons, blocked by the magnetic filter, prevents ionization and the reforming of new molecular ions. Successful magnetic filters were developed for the JET PINI source [5] and also for the "10×10" source developed at Lawrence Berkeley Laboratory by Pincosy [7]. However despite considerable experimentation, a model that could explain in detail how these sources operate was never achieved although a general understanding of the filter was achieved [6]. The only significant difference between the sources used for proton (or deuteron) enhancement and those where $H^-$ or $D^-$ production is required is the reversal of the accelerator polarity and design and the use of a slightly stronger magnetic filter (Holmes et al [8]) where the field is roughly doubled. Holmes [9] developed a model of the magnetic filter in its role in $H^-$ production, but no reference was made to positive ion species.

In view of the importance of the magnetic filter for the successful development of negative ion sources, it is essential that the theoretical basis underpinning the use of magnetic filter is improved. This paper addresses this issue with particular relevance to the enhancement for $H^+$ and $D^+$ where there is a considerable body of experimental data. All aspects of negative ion modelling are retained but will not be discussed at length here.

## 2. The Model Concept

This model is a major extension of the earlier one dimensional model of a negative ion source by Holmes [9], which divided the source chamber into a zero-dimensional driver region with fast electrons or primaries, responsible for ionization, a one dimensional plasma region where the fluxes of thermal electrons and ions is controlled by a set of transport coefficients and lastly a plasma grid boundary condition, this being the final boundary before the ions are extracted as a beam. Unfortunately this model did not provide a good basis for a filter model including ion species because, in reality, molecular ion species are not converted into other ions at the point of creation and it does not easily allow an accurate determination of the presence of atomic hydrogen. If these processes are ignored then the model is a reasonable argument for a negative ion source without surface production where the role and presence of fast atomic hydrogen is not critical.

To include the species evolution and atomic density calculation the model was modified to represent the primaries as a one dimensional fluid in the same form as the plasma electrons and ions. This allows the description of poorly confined sources such as the pancake source described by Ehlers and Leung [10], which forms an almost pure $H_2^+$ discharge where there is a large loss of primaries that are mobile throughout the entire source volume. An earlier version of the model described here was successfully used to analyse the data from that experiment [11] with the intention of creating a source that could simulate the effects of $D^+$ acceleration in an RFQ and subsequent drift tube linac for the IFMIF project for ITER.

*2.1 Global energy flow in the Discharge*
Before plasma transport is examined, the primary flux and energy flow needed to sustain the plasma is required. The plasma modelling is divided into three parts, the primary electron input, the plasma itself and finally the plasma sheath at the plasma grid, which normally floats at a negative potential due to the influx of primaries. In most sources there is zero current to this grid as the ion and total electron currents cancel. The code however retains the possibility of applying an external potential bias and positive currents are defined as net ion flow to the grid.

The plasma itself is biased by a potential, $\varphi$, positive with respect to the anode (ion source outer wall). The plasma electrons and the degraded primaries have two loss routes: to the anode and across the filter and each electron removes an energy $eT_1$ ($T_1$ is the driver plasma temperature) as it escapes. The model assumes a single loss route to the wall and this argument is justified by the existence of a potential difference across the filter region which causes the hot thermal electrons to lose almost all their energy, $eT_1$. This avoids having to implement a spatial variation of energy removal as the electron temperature decreases across the filter within the source. There is an energy balance for the main plasma where the driver thermal electron temperature is determined by the input energy to the thermal plasma derived from the energy transfer from the fast primary electrons. Current balance at the anode determines the plasma potential, $\varphi$ which determines the energy removed by the escaping plasma ions.

The ions escape with an energy $e\varphi$ if they are collected on the anode and a much larger value, $e(\varphi+V_g)$ if they are collected on the plasma grid. Note that $V_g$ is defined as a positive value when plasma grid is negatively biased with respect to the plasma. The net current to the plasma grid is a function of the combined energy distribution of the thermal and primary electrons. The potential energy, $V_g+\varphi$, lost by both groups of electrons in crossing the sheath in front of the plasma grid is transferred to the ions so they can escape with an equal energy. For this reason, the plasma grid net current is input data to the model and is used to determine the grid potential, $V_g$.

The primary electron power is $I_e(V-V_{fil})$ where $I_e$ is the primary emission current (close but not equal to the arc current) and is an input variable to the code, V is the arc voltage and $V_{fil}$ is the drop in voltage across the filament. For simplicity, we ignore the fact that the primaries are slightly more energetic by the plasma potential, $e\varphi$. This is justified as $\varphi$ is typically 2 to 5 volts and is very small compared with the arc voltage. This energy is either transferred to the plasma electrons or is lost through inelastic collisions using the value developed by Hiskes and Karo [12] (which includes ionization). The primaries then escape with just the thermal temperature, $T_1$, if degraded by inelastic collisions but a few (a fraction, $k_f$) escape with their full energy if they find a leakage path from the plasma. The two leakage areas are either the anode or the plasma grid but the exact leak rates are not known at the start of the calculation and are determined by iteration.

There is a final current balance at the grid. The primaries and plasma electrons are retarded by the grid potential, $V_g$, so that the algebraic sum of the electron and ion currents is equal to the grid current. The ion flux in turn is defined by the ionization rate less any losses to anode or filament and matches the ion flux surviving the transit across the plasma. However, again, the ion leaks are not known at the start of the calculation and are found by iteration. There can be a situation where an external amplifier or power supply is used to control the grid bias. In this instance, an extra power input of $I_gV_g$ is added, where $I_g$ can be positive or negative (the latter occurs when $V_g$ is less than the floating potential where $I_g$ is zero). This extra power is added to the ion energy flux at the grid. This is described in the section on the plasma grid.

**3. Model Methodology**

The plasma model for ion and electrons is very similar to that described by Holmes [9] and is shown in figure 1 in a flowchart format (for simplicity, not all elements of the model are included).

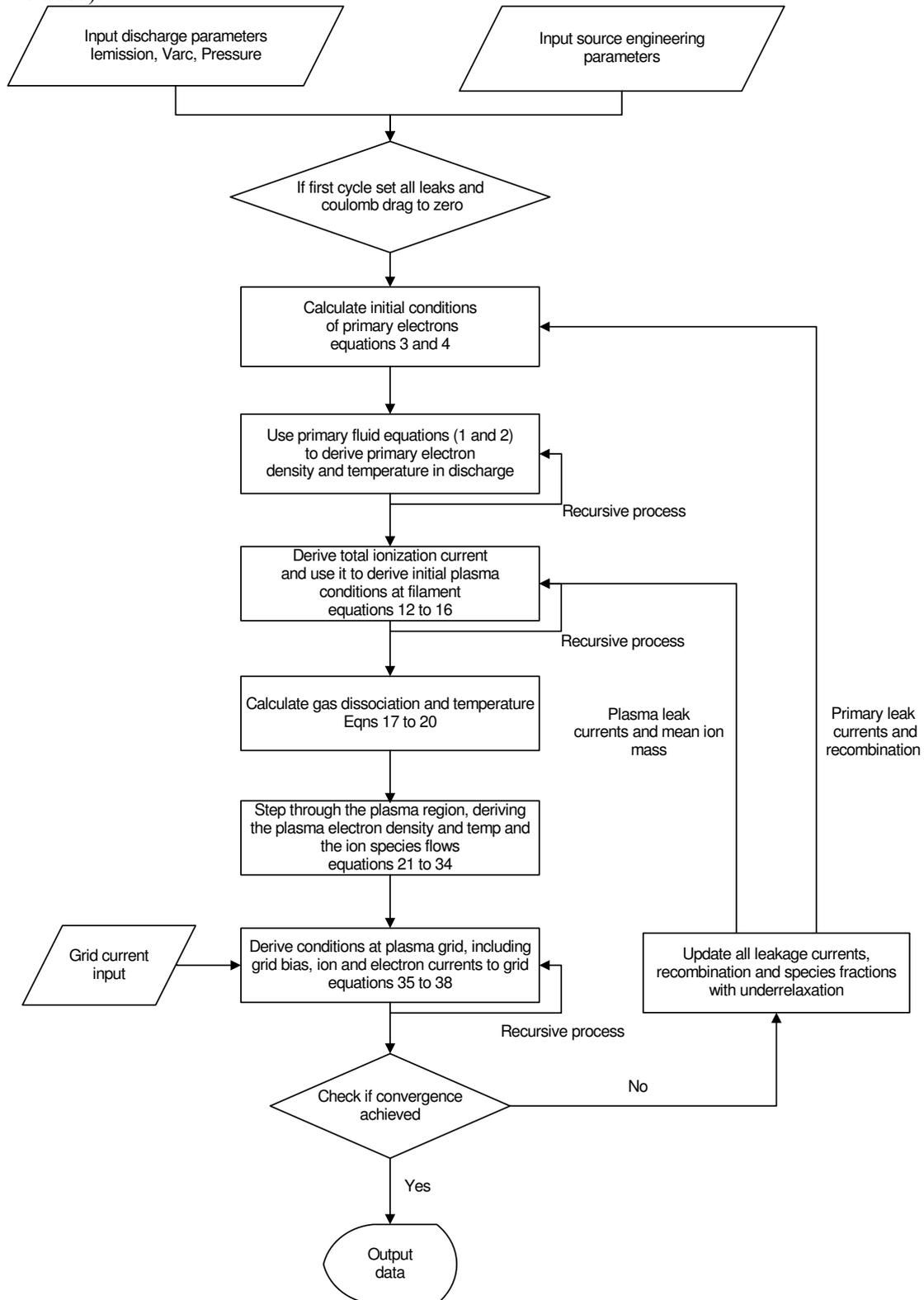

**Figure 1** Schematic flow diagram of the major sections of the model calculation.

The basic structure is to initially create a very simple model of the plasma without plasma leaks, coulomb drag or $H_3^+$ ions. This allows the formation of a preliminary primary electron distribution which is used to create a plasma distribution. Then first estimates of the leaks and coulomb drag are made together with the revised ion species distribution. This information, suitably relaxed to avoid numerical oscillation, is fed back into the primary distribution and the whole process is repeated until convergence is achieved. With this approach the model can incorporate almost all the physics known to exist inside a multipole source with as much detail as desired. All the collision processes are temperature dependent and this is included via suitable fitting functions, so that the local rate process depends on the local electron temperature. For convenience the data input is divided into discharge variables (i.e. arc voltage, current and pressure) and "engineering" variables which describe the source chamber (i.e. cusp field, filament position, plasma grid area etc.)

**4. The primary fluid equations**

The model begins by solving the plasma fluid equations for primary electrons developed by Epperlein and Haines [13] and modified for use in plasma sources by Holmes [9]. These are similar to those for plasma electrons described by Holmes [9] but the high primary temperature allows the electric field terms to be dropped and the collision frequencies differ from the plasma electrons. These primary electrons now have a local temperature and density and providing an initial estimate is made for the leak losses to the source electrodes, the primary distribution throughout the source volume can be derived from the fluid equations. The initial primary density and temperature are set by the energy and current balance at the filament itself. Once the spatial distribution is known, the total ionization current $I_i$ can be found from the variation of the ionization cross-section with electron energy. Most of these cross-sections can be found in the paper by Chan [14].

The fluid transport equations for the primary electrons are:

$$F_f = \frac{-eT_f \frac{dn_f}{dz}}{m\upsilon_{f1}\left(1+\left(\omega/\upsilon_{f1}\right)^2\right)} + \frac{en_f \frac{dT_f}{dz}}{2m\upsilon_{f2}\left(1+\left(\omega/\upsilon_{f2}\right)^2\right)} \quad (1)$$

and

$$Q_f = \frac{-eT_f^2 \frac{dn_f}{dz}}{m\upsilon_{f1}\left(1+\left(\omega/\upsilon_{f1}\right)^2\right)} - \frac{1.92en_f T_f \frac{dT_f}{dz}}{m\upsilon_{f2}\left(1+\left(\omega/\upsilon_{f2}\right)^2\right)} \quad (2)$$

The subscript f stands for fast (primary) electrons and the flux, $F_f$, is a particle flux, and $Q_f$ is an energy flux measured in eV/m²/sec. The term, m is the electron mass, and ω is the electron cyclotron frequency. Note that the temperature is in units of eV. The two collision frequencies are:

$$\upsilon_{f1} = NS_{el} + 2kn\lambda T_f^{-3/2}$$
$$\upsilon_{f2} = 2kn\lambda T_f^{-3/2} - 2NS_{el}$$
$$\omega^2 = \left(\frac{eB}{m}\right)^2$$

The term k is the coulomb collision rate coefficient of value k=7.7×10⁻¹² [(eV)^(3/2)m³s⁻¹] for 90 degree scattering of fast electrons from Huba [15], with a factor of two introduced to include scattering of the primary electrons from both ions and slow plasma electrons. The coulomb logarithm, λ, is approximately 10, N is the total gas density (molecular and atomic), n is the plasma density and $S_{el}$ is the electron elastic scattering rate given by Tawara et al [16] as

$$S_{el} = 6.1\times 10^{-14} \exp(-T_f) \quad [\text{m}^3\text{s}^{-1}]$$

Note that $\upsilon_{f2}$ is normally negative by convention but the charge, e, is defined as positive. The current density and energy flux, $F_f$, and $Q_f$ are positive for heat flowing from the source backplate (at z = 0) towards the grid and the density and temperature gradients are negative.

There is one modification to this simple picture which relates to the primary motion away from the electron emission site at the filaments towards the backplane of the source. It is assumed that the backplane acts as a perfect mirror for energy and particle fluxes so that there is no net flow (although there is a small leak of primary electrons, ignored for this purpose), so that $F_f$ and $Q_f$ are both zero and then $dn_f/dz$ and $dT_f/dz$ are also zero Thus in the region between filaments and backplane, the density and temperature are constant unless there is a spatially variable magnetic field present.

The effective source depth is the mechanical depth of the source, D, less the thickness of the cusp confinement on the backplane, $D_c$ (typically between 2cm and 5cm). Thus the total plasma depth, $D_s$ is:

$$D_s = D - D_c$$

The spatially varying section of region of the plasma, is primarily between the filaments and the extraction grid, is also described by (1) and (2). It is assumed that the filter lies mainly between the filaments and the plasma grid and is described by a gaussian field intensity distribution.

*4.1    Solution of the primary fluid equations*

In order to solve these equations some boundary conditions must be set. The initial primary temperature (eV) is mainly determined by the arc voltage, V, less the voltage drop across the filament, $V_{fil}$ and is:

$$T_{f0} = 2(V - V_{fil})/3 \tag{3}$$

The plasma potential is ignored here as it is small compared with the filament drop. The 2/3 factor arises from the kinetic theory of gases relating temperature to total energy and assumes that the ballistic nature of the primary electrons as they are emitted from the filament is modified into a distribution close to a Maxwellian. The primary electron flux, $n_{f0}v_f$, is found from the emission current, $I_e$, (this is *not* the cathode or arc current) less any electron emission current collected by the anode. The initial primary density is given by an equation similar to (11) in Holmes [9].

$$(I_e - k_f I_{fg})/e = k_f A_a n_{f0} v_f / 4 + (n_{f0} N R_{in} / T_f + 2 n_1 c_e T_f^{-3/2}) A_g D_s \tag{4}$$

where $I_{fg}$ is the fast electron current to the plasma grid of area $A_g$. Note that $I_{fg}$ is a positive quantity. The second term in the bracket is the coulomb drag loss, which heats the thermal density, $n_1$. The term $A_a$ is the anode cusp loss area and $R_{in}$ is the inelastic energy loss rate from Hiskes and Karo [12]. The parameter, $k_f$, represents the effectiveness of the lost primaries in imparting energy through inelastic collisions (i.e. if $k_f=1$ then these electrons do not have any inelastic processes). This rate, $R_{in}$, is:

$$R_{in} = 2.4 \times 10^{-12} \exp(-28/T_{f0}) \qquad \text{eV.m}^3\text{.sec}^{-1}$$

The rate, $R_{in}$, can be converted to a simple collision rate by dividing by the primary temperature. A simple expression for $k_f$ can be derived by comparing the characteristic time for primary electron loss to the walls with the inelastic loss collision time. If the former is much shorter than the latter, then $k_f$ approaches unity while the reverse leads to $k_f$ nearing zero. A suitable function would employ the exponential form, so that:

$$k_f = exp(-\nu \tau_w) = exp\left(-\frac{N R_{in} D_s}{v_{f0} T_{f0}}\right)$$

Hershkowitz et al [17] have argued that the plasma loss to the anode is equal to the total cusp length, C, multiplied by twice the hybrid Larmor diameter. This argument is extended to the energetic primary electrons, but as they are decoupled from the plasma electrons by their much higher energy, the primary electron loss area to the anode is twice the primary Larmor diameter multiplied by the total cusp length, C. Thus:

$$A_a = \frac{4Cm}{eB_c}\left(\frac{8eT_{f0}}{\pi m}\right)^{1/2}$$

Unfortunately the solution of (1) and (2) is very prone to instability, particularly as $F_f$ and $Q_f$ are both functions of the distance. A more practical method is to seek a solution that is consistent with (1) and 2 and retains a realistic attenuation of the current density and energy flux. Firstly (1) and (2) are rewritten in a more compact form where the subscripts have been dropped:

$$f = F_f = -aTn' + 0.5bnT' \tag{5}$$

$$q = -Q_f/T = aTn' + 1.92bnT' \tag{6}$$

The terms a and b are defined as:

$$a = \frac{e}{m\upsilon_{f1}(1+(\omega/\upsilon_{f1})^2)} \qquad b = \frac{e}{m\upsilon_{f2}(1+(\omega/\upsilon_{f2})^2)}$$

Note that f is a positive and q is a negative quantity and that n', T' are also negative. The term a is positive but b is negative if coulomb collisions are small due to the high value of the primary temperature. A solution is needed where n and T decrease in an assumed form of exponential decay with distance from the filament at z = 0. One possibility is:
Let

$$n = n_0 \exp\left(-\int \frac{\alpha}{a} dz\right) \tag{7}$$

and

$$T = T_0 \exp\left(-\int \frac{\beta}{b} dz\right) \tag{8}$$

After substituting (7) and (8) into (1) and (2) and applying the boundary conditions, the solution for the density and temperature is:

$$n_f = n_{f0} \exp\left(-\int \frac{0.234\upsilon_{f1}(1+(\omega/\upsilon_{f1})^2)}{\sqrt{eT_f/m}} dz\right) \tag{9}$$

and

$$T_f = T_{f0} \exp\left(\int \frac{0.33\upsilon_{f2}(1+(\omega/\upsilon_{f2})^2)}{\sqrt{eT_f/m}} dz\right) \tag{10}$$

Note that $\upsilon_{f2}$ is negative in (10), so both $n_f$ and $T_f$ decrease with distance from the start line (z = 0) as expected. The total ionization current, $I_i$, is:

$$I_i = \int_0^{ds} n_f(N_2 + x_0 N_1) q_i(T_f) dz - I_{rec} \tag{11}$$

where $N_2$ is the molecular gas density, $N_1$ is the atomic gas density, $q_i(T_f)$ is the *molecular* ionization rate (see section 6), $x_0$ is the multiplier for atomic ionization ($x_0 \approx 0.66 \pm 0.04$) and $I_{rec}$ is the ionization current lost by recombination of $H_3^+$ with H to reform $H_2$ molecules (see section 8.1). Note that the atomic ionization rate is almost exactly 0.66 of the molecular rate at all energies slightly above the threshold, so for convenience, only the molecular rate is used. This $I_{rec}$ current is not known in the first cycle but can be derived from the build up of the ion species described in section 8.1.

*4.2    Numerical solution*
The primary effectiveness, $k_f$ is initially set to zero and hence a value for $n_{f0}$ can be derived from (4) and the arc voltage gives a value for $T_{f0}$. This latter value does not change in later cycles. As $T_f$ appears in the argument of the exponent in (10), care is required for stepwise integration because of the changing value of $\omega$, the cyclotron frequency. Differentiating (10) gives:

$$T_f' = T_f \frac{0.3298\upsilon_{f2}(1+(\omega/\upsilon_{f2})^2)}{\sqrt{eT_f/m}}$$

The derivative of $T_f$ varies with the local square root of $T_f$, which can be integrated (where $\omega$ is a local constant) to get:

$$T_{fn} = \left[\frac{0.3298\upsilon_{f2}(1+(\omega/\upsilon_{f2})^2)}{2\sqrt{e/m}}\delta + \sqrt{T_{fn-1}}\right]^2$$

where $\delta$ is the local increment along the z axis from the n-1 position. Once the new value of $T_f$ is obtained, an average root value is made over the step and used in (9) to obtain the change in $n_f$. At the same time the local ionization current, $I_i$, is obtained by multiplying by the ionization rate and the density of atomic and molecular gas.

Once a solution is found for the plasma density in the driver region, the coulomb drag element can be included as a correction in subsequent cycles and a new value of the coefficient, $k_f$, can be assigned.

**5. Driver plasma equations**
The driver plasma model is similar to that of Holmes [9] but somewhat simpler as the primary fluid equations of Section 4 replace the original description of the primary electrons. Firstly there is ion continuity:

$$I_i - I_{ig} = n_1 e \psi_i (A_{anode} T_1 + A_f T_1^{1/2}) \tag{11}$$

where

$$I_{ig} = f_i I_i$$

Here $I_{ig}$ is the plasma grid ion current loss defined by a fraction, $f_i$, of the total ionization current and $T_1$ is the electron temperature at the filament (z=0). The area, $A_f$, is the filament area for ion loss and $A_{anode}$ is the anode loss area for plasma electrons and ions defined by four fold the hybrid Larmor radius [17] at the anode surface multiplied by the total cusp length, C:

$$A_{anode} = \frac{4C}{eB_c}(4a_{mass} m_p \psi_i (1+\eta)^{1/2} m\psi_e)^{1/2}$$

with

$$\psi_i = 0.6\left(\frac{e}{a_{mass} m_p}\right)^{1/2} \qquad \psi_e = \frac{1}{4}\left(\frac{8e}{\pi m}\right)^{1/2} \qquad \text{and} \quad \eta = \phi/T_1.$$

where $m_p$ is the proton mass and $m$ is the electron mass. The terms $\psi_i$ and $\psi_e$ are convenient as they allow the effects of $T_1$ to be extracted.

The total ionization current (protons and molecular ions) is $I_i$ and is found from the primary electron profile and the atomic and molecular density as described in section 4.1. However the plasma grid ion current, $I_{ig}$ is not yet known and is treated as a correction to be updated when the plasma grid is analysed in section 8.2.

A similar equation for electron continuity can be written:

$$I_i - I_{eg} + I_e - I_{fg} - I_{fa} = eA_{anode} n_1 \psi_e T_1 exp(-\eta) \tag{12}$$

where

$$I_{ig} = f_i I_i$$
$$I_{eg} = f_e f_i I_i$$

The value of $f_i$, $f_e$ and $I_{fg}$ are found from the current balance at the plasma grid and are calculated in later cycles but an initial estimate is required. The primary current to the anode, $I_{fa}$, is derived below. It is assumed that the primaries that are not lost at the anode or plasma grid become thermal electrons and add to the net thermal electrons formed in the total ionization (less those lost directly to the plasma grid). No electrons are lost on the filament.

The $k_f$ term is not used in (12) as this equation is concerned with charge flow and not the efficiency of use of primary electrons. The plasma electrons that go to the anode cross a retarding sheath potential, $\varphi$, where $\eta = \varphi/T_1$. The arc current (or anode current as it normally defined), $I_a$, is:

$$I_a = I_e - I_{fg} + I_i - I_{eg} - I_{ai} \tag{13}$$

Finally there is the energy balance equation for the thermal electrons. The volumetric input energy is the same as in the model in Holmes [9] except $k_f$ is now included as the efficiency of energy transfer from primaries is involved:

$$W = \frac{\left(I_e - k_f[I_{fg} + I_{fa}]\right) n_1 k \left[V_{eff}^{0.5} - 2T_1^{0.5} - T_1/V_{eff}^{0.5}\right]}{NR_{in} + 2n_1 k/T_f^{0.5}} \tag{14}$$

For simplicity the $T_f$ and $n_f$ values are those at the backplate (z = 0). This energy is fed exclusively to the electrons and ions created by ionization and the $k_f$ term is applied to the lost primaries as (14) is an energy balance. The retarded primaries are assumed to retain an energy, $eT_1$, as they escape as they are identical to plasma electrons. Thus applying this energy, W, to the thermal electrons yields:

$$W = I_i T_1 (1 + \eta) \tag{15}$$

Equations (11) to (15) can be solved by iteration to obtain values of $n_1$, $\eta$, and $T_1$. This allows a coulomb loss term to be inserted in the equations for the two collision frequencies governing the primary flow as there is an initial guess for the plasma coulomb drag on the primary electrons based on the first cycle calculation.

Again the there are primary loss terms which are not yet defined; namely $I_{fg}$ the primary loss to the grid and $I_{fa}$, the loss to the anode. The latter can be expressed by the equation:

$$I_{fa} = en_f y_f \left(\frac{C}{eB_c} m v_f\right) = \frac{C}{B_c} \frac{8 n_f T_f}{\pi}$$

Thus the first stage of the code is to cycle through the primary model and the driver plasma until a solution is found. However the exact flux of ions, primaries and plasma electrons to the plasma grid are not known, and similarly for the ion species fractions at the plasma grid and the plasma density at the extraction plane. Once these fluxes are known, the mean ion mass and the primary and ion currents to the extraction plane can be updated as this depends on the actual total grid current. There is also a loss of plasma electrons to the grid which will be updated at the same time.

## 6. Modelling the Neutral Gas

*6.1 Derivation of the atomic and molecular fractions*
The next step is to derive the atomic gas fraction. For simplicity no negative ion rates are included as these are not expected to impact significantly on the atomic gas fraction. For hydrogen, the ionization rate for atomic hydrogen is 0.66 of the molecular rate at all energies within an error of ±5% (Chan [14]). The atomic fraction is assumed to be constant everywhere inside the source, so the calculation is global for the entire source chamber. This is particularly true for atomic loss on the walls, which is obviously a global process. The complicating factor is the contributions made to protons and atomic hydrogen by $H_3^+$ dissociation and recombination. Unlike the other processes these are local and mainly take place away from the filament region.

To circumvent this problem, these processes will be treated as perturbations that are initially zero and introduced as additional source or sink terms in later cycles. The processes that contribute are from Chan [14] and Green [18] except the ionization rate which from Freeman and Jones [19].

The rates are introduced as an algebraic empirical expression that fits the experimental data over the temperature range up to about 100eV for ionization and other energetic collisions. For low energy collision processes it is valid up to about 20eV. This avoids problems with polynomial fits which diverge if the solution process goes out of range. The only exception is the xs2 rate which is fitted by two different expressions above or below $T_f = 22$eV.

**Table 1.** Analytic forms of reaction rates used in the model.

| Process | label | Expression in $m^3.sec^{-1}$ |
|---|---|---|
| $H_2 + e \rightarrow 2H + e$ | xs2 | xs2=$10^{-15}$exp(3.2)exp(-8/T)F |
| | | if T>22, then F = 1 |
| | | else F = exp(-0.068(T-22)) |
| $H_2^+ + e \rightarrow 2H$ | xs4 | xs4=$10^{-15}$exp(3.5)(1-exp-(5.2/T)) |
| $H_2^+ + e \rightarrow H^+ + H + e$ | xs5 | xs5=$10^{-15}$exp(4.5)exp(-2.5/T) |
| $H_2^+ + H_2 \rightarrow H_3^+ + H$ | xs6 | $2.1 \times 10^{-15}$ |
| $H_3^+ + e \rightarrow H^+ + H_2 + e$ | xs7 | xs7 = $10^{-15}$exp(6.5)exp(-13/T) |
| $H_3^+ + e \rightarrow H_2 + H$ | xs8 | xs8 = $10^{-15}$exp(3.8)(1-exp(-6/T))) |
| $H_2 + e \rightarrow H_2^+ + 2e$ | $q_i$ | $q_i$=$10^{-15}$exp(3.98)exp(-22.84/T) |

Chan et al [20] has derived a particle balance equation that can be used to derive the atomic density, $N_1$. In the first order any process that contains $H_3^+$ is ignored to obtain:

$$\frac{N_1 \gamma}{t_1} + 0.66 N_1 n_{f0}^* q_i = 2 N_2 \left( n_e^* xs2_e + n_f^* xs2_f \right) + n_2^* n_e^* \left( xs5 + 2 \times xs4 \right)$$
$$+ N_2 n_2^* xs6 + n_3^* n_e^* xs8$$

Here $N_2$ is the gas density, $n_2$ is the $H_2^+$ ion density and $n_e$ is the plasma density. The term $\gamma$ is the accommodation coefficient on the walls (Chan [10] selected a value of 0.3 for $\gamma$). The term, $t_1$, is the transit time of an atom in the source (=$4D/v_{atom}$). The asterisk indicates that an average is taken of the densities across the source depth (i.e. initial/2 + final/2) to allow for the spatial distribution.

In addition we have the condition that the total mass flow, $Q_g$, of gas through the source is constant, so:

$$Q_g = 2m_p N_0 \kappa T_w^{1/2} = 2m_p N_2 \kappa T_g^{1/2} + m_p N_1 \sqrt{2} \kappa T_g^{1/2} \quad (16)$$

Here the initial gas input is at the wall temperature, denoted by $N_0$ and $T_w$ but emerges into the accelerator as hot gas of temperature, $T_g$, with densities $N_1$ and $N_2$. The term $\kappa$ is the fixed part of the gas conductance and includes the number and area of the extraction apertures.
If $N_1/N_2 = G$ then rationalization gives:

$$\sqrt{\frac{T_w}{T_g}} N_0 = N_2 \left( 1 + G/\sqrt{2} \right) = N_c \quad (17)$$

Here $N_c$ is the gas density in the presence of the discharge (i.e. the gas is hot) but without dissociation. Note that $N_1/2 + N_2$ is not equal to $N_c$ because of the increased mobility of the atomic gas.

Each of the ion densities is related to the plasma density, $n_e$, by a coefficient, $c_{n1}$, $c_{n2}$ or $c_{n3}$:

$$n_1 = n_e c_{n1}$$
$$n_2 = n_e c_{n2}$$
$$n_3 = n_e c_{n3}$$

The values of the molecular ion fractions, $c_{n2}$ and $c_{n3}$, are not known initially so a preliminary guess of 0.333 is assumed but this is updated after each major cycle of convergence.

An ion species model – model descriptionThe values of $n_e$, $T_{f0}$ and $T_e$ are found from the solution of the driver region giving all rates. The atomic gas velocity is derived from an assumed initial gas temperature of 500K. The initial value of $n_{f0}$ is known and the undisturbed gas density, $N_0$, is obtained from the initial gas temperature. Simplifying the particle balance equations developed by Chan et al [20] using the coefficients below:

$$R = \frac{\gamma}{t_1} + 0.66 n_{f0}^* q_i \quad H = 2\left(n_e^* xs2_e + n_{f0}^* xs2_f\right) \quad B = c_{n2} n_e^{*2}(xs5 + 2xs4)$$

$$C = c_{n2} n_e^* xs6 \qquad\qquad D = c_{n3} n_e^{*2} xs8$$

The balance equation for molecular gas is then:
$$GN_2 R = HN_2 + B + CN_2 + D$$

Substituting $N_2$ with $N_c$ using (17) gives after rationalization:
$$G = \frac{H + C + (B+D)/N_c}{R - (B+D)/(\sqrt{2}N_c)} \tag{18}$$

Once G is known, $N_2$ can be found and hence:
$$N_1 = N_c / (G^{-1} + 1/2^{0.5}) \tag{19}$$

*6.2 Gas Temperature*

The ions are heated by energy transfer from the electrons by coulomb collisions. Hence using the driver region conditions for simplicity:
$$eA_w \gamma_h (T_g - T_w) N v_{gas} / 4 = Q_i = k_\varepsilon e A_g d_s \lambda n_1^2 T_1^{-1/2} / a_i$$

Here $T_w$ is the wall temperature in eV, $T_g$ is the hot gas temperature, $k_\varepsilon = 3.2 \times 10^{-15}$ [eV]$^{3/2}$m$^3$s$^{-1}$ is the coulomb rate coefficient for energy equilibration between ions and electrons from NRL [15], $\lambda$ is the coulomb logarithm (= 10), $A_w$ is the total gas cooling surface area, $A_g d_s$ is the grid area times plasma depth and lastly $a_i$ is the ionic mass number (= 1 for protons). The term $\gamma_h$ is the thermal accommodation coefficient and is probably similar to the atomic recombination coefficient and $v_{gas}$ is the gas velocity. However not all the ions have charge exchanging collisions before hitting the anode so an additional reduction of $1 - \exp(-N\sigma d_s)$ is included in the numerator on the RHS. The product $Nv_{gas}$ is conserved as the gas flow is constant and this simplifies to:

$$T_g = T_w + \frac{4 A_g d_s k_\varepsilon n_1^2 T_1^{-1/2}[1 - \exp(-N\sigma d_s)]}{A_w a_i \gamma_h N_0 \left(\frac{2.55 e T_w}{M_g}\right)^{1/2}} \tag{20}$$

The gas density scales as $(T_w/T_g)^{1/2}$ as the gas mass flow is constant as shown in (17). This is used to update the value of $k_f$ at the end of a cycle.

*6.3 The gas coefficients $\gamma$ and $\gamma_h$*

In the above analysis two wall coefficients have been used; $\gamma$, the atomic hydrogen recombination rate and $\gamma_h$, the thermal accommodation rate. In the work by Chan et al [19], only the first was used and was treated as a fixed constant. The values are not known with any certitude and can only be determined by comparing the code results with experimental results. However Chan, et al argue that $\gamma$ should be about 0.2 for deuterium discharges although experimental data from Wood and Wise [21] suggest it should a little smaller. In the results presented here the values given in table 2 are used.

The values of $\gamma$ are closer to the results of Wood and Wise [21] but it should be emphasised that the presence of the plasma and the evaporated tungsten on the copper walls makes it difficult to have a definitive value. A sensitivity study varying these parameters over the range 0.1 to 0.4 showed a modest influence on the species ratios for both with the stronger effect being a decrease in deuteron yield with increasing $\gamma$ but an increase in yield with $\gamma_h$.

**Table 2.** Values of the number and energy recombination coefficients used for modelling deuterium.

|  | $\gamma$ | $\gamma_h$ |
|---|---|---|
| Jet PINI Source | 0.13 | 0.35 |
| LBL Source | 0.35 | 0.35 |

## 7. The Negative ion density

Although not strictly part of the processes involving the formation of protons (or deuterons) the role of negative ions in the plasma must be included for two reasons. Firstly the difference between a filter source designed to yield negative ions and one creating a high yield of protons is fairly small, a mere factor of two in the filter field strength, so the processes are highly linked. This can be seen more clearly in the following sections discussing the motion of electrons through the filter where these electrons spend part of their existence as negative ions, which alters their mobility and other transport coefficients. Secondly the presence of negative ions at the extraction plane after the filter fields alters the boundary conditions at that plane. Thus negative ions formed by volume processes as discussed below, must be included in the total model even for applications to positive ion production. For sources designed to optimise the production of negative ions, surface production is used to enhance the density but this will not be discussed here.

The negative ions are formed by dissociative attachment collisions between vibrationally excited hydrogen molecules and cold electrons whose temperature is less than 2eV [22, 23]. Above this temperature, the rate coefficient decreases rapidly and at the same time the electron detachment rate rises significantly, sharply reducing negative ion formation. The vibrationally excited molecules are formed by collisions between fast electrons (primaries) and molecules in the lowest vibrational state ($\upsilon = 0$) and are destroyed by wall collisions or electronic excitation or ionization. The analytical forms of the reaction rates are given in table 3.

**Table 3.** Analytic forms of reaction rates used in the model for negative ions.

| Process | Label | Value ($m^3 s^{-1}$) |
|---|---|---|
| $e + H_- = H + 2e$ | $S_{EV}$ | $1.5 \times 10^{-14} T_e^{2.5}$ |
| $e_{fast} + H_- = H + e + e_{fast}$ | $S_{FV}$ | $9 \times 10^{-13}$ |
| $H_- + H = 2H + e$ | $S_H$ | $1 \times 10^{-15}$ |
| $e + H_2(\upsilon > 8) = H_- + H$ | $S_{DA}$ | $3 \times 10^{-14}/(1+T_e^{1.5})$ |
| $H_+ + H_- = 2H$ | $S_{II}$ | $5 \times 10^{-14}$ |

As the negative ions cannot move far from where they are created without being destroyed, the mean free path is typically a few centimetres, their number density can be described by a pair of rate equations. The first of these equations describes the vibrational density which is uniform throughout the plasma chamber:

$$N_{(\upsilon \geq 8)} n_f \frac{R_{in}}{T_f} + \frac{N_{(\upsilon \geq 8)}}{\tau_w} = N n_f S_P \tag{21}$$

Only vibrational levels in excess of 8 are included as the lower levels have a dissociative attachment rate that is roughly a factor of 5 lower for each level below 8. This is reflected in the value of $S_p$, the rate coefficient for production from levels with $\upsilon \geq 8$. The value of the wall loss time, $\tau_w$, depends on three factors: the gas temperature derived in Section 6, the size of the source and lastly the number of wall collisions before de-excitation. Hiskes and Karo [24] have argued that the collision number should be approximately 4. The destruction rate is the same as the total inelastic collision rate, $R_{in}$, developed by Hiskes and Karo [12] and used in Section 4.1.

A similar rate balance equation can be written for the negative ions (Holmes [9]). The production rate, $S_{DA}$, by dissociative attachment with vibrationally excited molecules above $v = 8$ is balanced by losses by ion-ion recombination with a rate, $S_{II}$, electron detachment with a rate, $S_{EV}$, and $S_{FV}$ for the primary electrons and also loss by atomic gas collisions with a rate, $S_H$. Wall losses are neglected as it is assumed that there is a dense plasma and hence small mean free path. This gives a balance equation:

$$N_{(v \geq 8)} n_e S_{DA} = n_- n_e S_{EV} + n_- N_1 S_H + n_- n_+ S_{II} + n_- n_f S_{FV} \tag{22}$$

Replacing the negative ion density, $n_-$, by the fractional negative ion density, $q = n_-/n_+$, and assuming local plasma neutrality, gives a quadratic equation in q:

$$0 = -n_+ S_{EV} q^2 + q\left(n_+ [S_{II} + S_{EV}] + N_1 S_H + n_f S_{FV} + N_{(v \geq 8)} S_{DA}\right) - N_{(v \geq 8)} S_{DA}$$

If the local value of the plasma electron temperature, $T_e$, $n_+$ and $n_f$ as well as the value of the vibrational density from (21) are all known, the value of q can be found, so deriving the electron and negative ion densities. As electrons now spend part of their time as negative ions, and during this time are virtually unaffected by the magnetic field. The term, q, can be used as a weighting term for this effect.

**8. Plasma electron and ion transport**
The sections above allow a description of the initial conditions of the plasma, including a preliminary derivation of the gas dissociation and temperature as well as the total ionization and primary electron distribution. All of this is based on the preliminary assumption of no leaks of electrons or ions to the grid and the non-existence of $H_3^+$. The next step is to derive the plasma electron distribution and a better estimate of the ion species distribution.

The basic equations for plasma transport are identical to those described by Holmes [9], so the expressions are:

$$F_e = \frac{-e n_e E - e T_e \frac{dn_e}{dz}}{m v_1 \left(1 + (\omega/v_1)^2\right)} + \frac{e n_e \frac{dT_e}{dz}}{2 m v_2 \left(1 + (\omega/v_2)^2\right)} \tag{23}$$

$$Q_e = \frac{-e n_e T_e E - e T_e^2 \frac{dn_e}{dz}}{m v_1 \left(1 + (\omega/v_1)^2\right)} - \frac{1.92 e n_e T_e \frac{dT_e}{dz}}{m v_2 \left(1 + (\omega/v_2)^2\right)} \tag{24}$$

$$F_i = \frac{e n_i E - e T_i \frac{dn_e}{dz}}{M v_i \left(1 + (\Omega/v_i)^2\right)} = \frac{n_i E - T_i \frac{dn_e}{dz}}{A_i} \tag{25}$$

In (23) to (25), $F_e$ and $F_i$ are the directed current densities of thermal electrons or ions towards the extraction plane and are positive if the flow is in that direction. The term, $Q_e$ is the energy flow [eVm$^{-2}$s$^{-1}$]. As before ion energy transport is neglected and the ion temperature gradient is ignored and the ion temperature $T_i$ is set equal to $T_g$ due to the high ion-gas collision rate.

The negative ions are not transported as negative ions but their presence is accounted for by creating weighted transport coefficients, using the negative ion fraction, q. At each position along the axis of the source chamber, the local magnetic filter field is evaluated and the collision rates determined from the electron temperature and density of the previous position (except for the first step position which uses the values $n_1$ and $T_1$). For the negatively charged particles these rates are:

Electrons
$$v_{e1} = N S_{el} + 2 k n_e \lambda T_e^{-3/2}$$
$$v_{e2} = 2 k n_e \lambda T_e^{-3/2} - 2 N S_{el}$$
$$\omega = eB/m$$

Negative ions
$$\upsilon_{n1} = NS_{ns} + n_e k_n \lambda T_e^{-3/2}$$
$$\upsilon_{n2} = n_e k_n \lambda T_e^{-3/2} - 2NS_{ns}$$
$$\Omega = eB/M_n$$

The value of the electron–gas scattering rate, $S_{el}$, is now given by [16]:
$$S_{el} = 6.1 \times 10^{-14} \exp(-T_e)$$

There is a factor of 2 before the Coulomb coefficient k in the above expressions to reflect that 90 degree scattering occurs via electron and ion impacts to cross the field lines just as in the case of the primary electrons in section 4. In the case of the negative ions, the coulomb scattering is smaller as it is purely ionic and Huba [15] gives the value for $k_n$ of $1.4 \times 10^{-13}$ [(eV)$^{3/2}$m$^3$s$^{-1}$] for equal mass ions. Smith and Glasser [25] suggest that a value for the negative ion- gas scattering rate, $S_{ns}$, would be $S_{ns} \approx 6 \times 10^{-16}$ [m$^3$s$^{-1}$] and this is assumed not to be temperature dependent. The composite transport coefficient is then:

$$m\upsilon_1 \left(1 + \left(\omega/\upsilon_1\right)^2\right) = (1-q)m\upsilon_{e1}\left(1 + \left(\omega/\upsilon_{e1}\right)^2\right) + qM_n\upsilon_{n1}\left(1 + \left(\Omega/\upsilon_{n1}\right)^2\right)$$

The other denominator is determined similarly. This argues that the electrons are serially electrons or negative ions and the fractional density, q, reflects this, so the final transport coefficient is the simple weighted sum of the two values. Note that the value of q is updated at each step and is almost zero in the driver region.

The ion transport coefficient is similarly treated as there are three ion species. The two terms which determine the value of $A_i$ for an ion of mass, M, are:

$$\nu_i = Nxs6\left(\frac{M_2}{M}\right)^{0.5} + k_n n_n \lambda T_e^{-3/2} \qquad \Omega = eB/M$$

where $M_2$ is the diatomic ionr mass. The presence of the mass ratio term in the gas collision rate reflects the higher velocity of the lighter positive ions and it has been assumed that the dominant scattering rate is the $H_2^+$ - $H_2$ collision process leading to $H_3^+$ and that this rate can be applied to the other species for the purposes of ion-gas scattering. It has been assumed that the positive ions only have effective scattering coulomb collisions with the negative ions.

The presence of the different ion species is included by creating a composite value of the $A_i$ term. If the fractional current density of each ion species is $C_1$ to $C_3$, then:

$$A_i = C_1 A_1 + C_2 A_2 + C_3 A_3$$

Here $C_1$, $C_2$, and $C_3$ are the fractions of the plasma grid ion current density in the three species and are not the same as the $c_{n1}$, $c_{n2}$ and $c_{n3}$ terms used earlier in (18), although they are related. The values of $A_1$, $A_2$, and $A_3$ do contain a magnetic element through the $\Omega$ term but its effect is almost negligible. The values of $C_1$, $C_2$ and $C_3$ are constant in any particular cycle of calculation and are only updated when the ion flux reaches the plasma grid, that is to say once per cycle following the build up of each ion flux as shown in the next section. This approach avoids instability problems in the code.

*8.1    Build up of the ion species fluxes*

The directed initial current density of ions and electrons towards the plasma grid are zero at the plane of origin at the back of the source as no ionization has yet taken place. By analogy, the directed electron energy flux is also zero. The ionic grid current is defined as a fraction, $f_i$, of the total ion production and the electron current is a further fraction, $f_e$ of this current. The former has an initial value close to unity while the latter is close to zero.

In the equations below the diamond bracket indicates a simple average of the flux product taken over the previous step (the first step assumes that values are zero). The velocities in the denominator are the mass weighted thermal ion velocities based on the gas temperature which is also the effective ion temperature due to the high ion-neutral collisionality. The term $x_0$ is the ratio of ionization rates for atomic and molecular gas and is 0.66. After each step the total fluxes of all ion species are re-evaluated by integrating the following expressions from the source backplate to the plasma grid:

$$\frac{dF_1}{dz} = n_f N_1 x_0 q_i + \frac{\langle F_2 n_e \rangle xs5}{v_2} + \frac{\{\langle F_3 n_e \rangle xs7_e + \langle F_3 n_f \rangle xs7_f\}}{v_3} \quad (26)$$

$$\frac{dF_2}{dz} = n_f N_2 q_i - \frac{\langle F_2 n_e \rangle xs5}{v_2} - N_2 \langle F_2 \rangle xs6 \quad (27)$$

$$\frac{dF_3}{dz} = N_2 \langle F_2 \rangle xs6 - \frac{\{\langle F_3 n_e \rangle xs7_e + \langle F_3 n_f \rangle xs7_f\}}{v_3} - \frac{\langle F_3 n_e \rangle xs8}{2v_3} \quad (28)$$

After integration from the backplate to the plasma grid, the total directed ion flux, $F_{ig}$, is composed of the three individual fluxes, $F_{1g}$, $F_{2g}$ and $F_{3g}$:

$$F_{ig} = (F_{1g} + F_{2g} + F_{3g}) \quad (29)$$

and:   $C_1 = F_{1g} / F_{ig}$   $C_2 = F_{2g} / F_{ig}$   $C_3 = F_{3g} / F_{ig}$

This integration is made after the electron spatial distribution has been found using (23) to (25) which is based on the ion current density fractions, $C_1$, $C_2$ and $C_3$, derived in an earlier cycle. The new integration allows an update of these fractions to be made. The values of the plasma density fractions, $c_{n1}, c_{n2}$ and $c_{n3}$ are derived by dividing the current density fractions by the normalised velocities appropriate to that ion.

The fraction of the total ion current that goes to the plasma grid is only part of the total ion production, typically ~60%, with the rest going to the anode or the filament structure. The addition of a magnetic filter increases the latter currents significantly and reduces the former current. These new fractions are used in the entire transport calculations (25) to define that average ionic transport coefficient for the cycle that follows the definition of the $C_1$, $C_2$ and $C_3$ values.

In the case of electron flow, it is assumed that the electron flux, $F_e$, is a fraction, $f_e$, of the total ion flux that is collected by the plasma grid. Hence:

$$F_e = f_e I_{ig} = f_e f_i e (F_{1g} + F_{2g} + F_{3g})$$

where

$$I_g = f_i I_i (1 - f_e)$$

The electron current to the plasma grid is a low fraction of the total electron production in almost all situations. The values of $f_e$ and $f_i$ are found by iteration over many cycles so that it matches the assumptions about the plasma grid current, $I_g$, that form part of the input data. This is discussed further in section 8.2. However it is necessary to guess the values initially but these are updated in later cycles.

If all the three ion flux gradients (26) to (28) are added, all the terms cancel apart from ionization (the $q_i$ terms) and the xs8 term in the gradient of $F_3$. This last term is the recombination of $H_3^+$ and $H_2$ and represents a loss of ionization. Thus:

$$I_{rec} = A_g \int_0^{ds} \frac{e \langle F_3 n_e \rangle xs8}{2v_3} dz \quad (30)$$

This current is included in the total ion current balance but is usually quite small.

*8.2 The Grid Boundary condition*
The random ion current density controlled by the ion sound speed that reaches the plasma grid of area $A_g$ must equal the directed ion current density that is transported via (25). This can be expressed by the equality:

$$ef_i F_{ig} = I_{ig} / A_g = en_{ig} \psi_i T_{eg}^{1/2} \tag{31}$$

Here $n_{ig}$ is the ion density at the grid following a solution of the equations in the transport Section 8 and $T_{eg}$ is the matching temperature. Equation (31) creates a new value of the ion fraction, $f_i$ based on the value of $n_{ig}$ and $T_{eg}$ and the value of $F_{ig}$ derived in section 8.1. The value of $f_i$ ,used in the next cycle, is created by mixing the original estimate with the new value derived from (31) above. Typically between 5 and 10 percent of the new solution is used, giving an under-relaxation parameter of around 0.1.

The other quantities to be derived are the plasma grid bias voltage, $V_g$ and the value of $f_e$. The former is found from current continuity:

$$A_g e \left[ \psi_e n_{fg} T_{fg}^{1/2} \exp\left(-\frac{V_g}{T_{fg}}\right) + \psi_n n_{eg} T_{eg}^{1/2} \exp\left(-\frac{V_g}{T_{eg}}\right) \right] = I_{eg} + I_{fg} = I_{ig} - I_g \tag{32}$$

The negative ions are assumed to behave as thermal electrons with the proton mass at the sheath, hence the composite value of $\psi_n$ for electrons and negative ions can be derived as:

$$\psi_n = \frac{1}{4}\left(\frac{8e}{\pi m}\right)^{1/2}(1-q) + \frac{1}{4}\left(\frac{8e}{\pi M_n}\right)^{1/2}(q)$$

The actual grid current is $I_g$, defined as positive if flowing conventionally from the grid (i.e. very negative bias of the plasma grid creates a positive value for $I_g$). This last equation is highly non-linear and must be solved using a step-wise procedure. The values of $n_{fg}$ and $T_{fg}$ are derived from the last primary electron distribution. If the filter is moderately strong there is no primary current. There is no explicit negative ion current at the plasma grid used in (32) as the negative ion velocity is small relative to that of the ions, but their presence is allowed for through the fact that the electron density is smaller than that of the positive ions.

The value of $V_g$ is used to derive an update for the electron fluxes via the expression:

$$f_e I_{ig} / A_g = e \psi_e n_{eg} T_{eg}^{1/2} \exp\left(-\frac{V_g}{T_{eg}}\right) \tag{33}$$

$$I_{fg} = e A_g \psi_e n_{fg} T_{fg}^{1/2} \exp\left(-\frac{V_g}{T_{fg}}\right) \tag{34}$$

Again the actual negative ion current density is ignored as it will be small compared with that of the electrons. Similarly to the convergence of the value of $f_i$, severe under-relaxation is needed for $f_e$. These new values of $f_i$, $f_e$ and $I_{fg}$ are returned to the primary modelling described in sections 4. and 4.1 together with the recombination current given by (30). Typically it takes about one thousand cycles to converge to better than 1% although the convergence rate depends strongly on gas pressure.

**9. Comparison of the Model with Experimental Results**
Two plasma generators, for which extensive experimental databases exist, have been considered. The first of these is the "10×10" source developed by Pincosy [7] at the Lawrence Berkeley Laboratory for neutral beam heating. Ion species measurements with and without a magnetic filter exist for this source together with electron temperature and density measurements orthogonal to the magnetic filter and parallel to the beam axis. The second source is the "PINI" source, used for neutral beam heating [1, 5, 8] on the JET experiment, for which extensive ion species measurements with and without a filter field exist. There are also positive ion species measurements when the source has a filter strong enough to allow it to act as a negative ion source, hence bridging the gap between the two modes of operation.

There are two different types of magnetic filter - the supercusp and dipole - as shown in figure 2. The JET "PINI" source has a supercusp type of filter created by the magnets outside the source chamber as shown in figure 2(a). The "10x10" source has a dipole filter created by rows of small bar magnets within the plasma generator that form a sheet of magnetic field instead of the external magnets shown in figure 2(b). In order to emphasize the filter cusp line in the PINI source the normal cusp confinement is of the checkerboard geometry where the north and south poles vary in two dimensions. Although the "10x10" source could have had the same arrangement, the decision to use internal bar magnets to form the dipole filter allows the normal cusp confinement to be of the linear type, where the polar alternation is in one dimension.

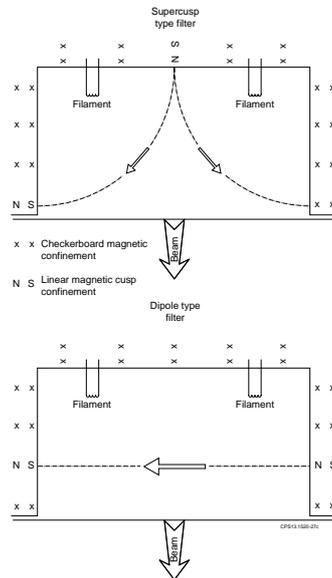

**Figure 2** The two types of magnetic filter field configuration: *upper* the supercusp field giving a long range three dimensional and *lower* the dipole field giving a one dimensional field.

In both cases the source filaments are located towards the back plane of the source and behind the filter field. The supercusp filter was developed to avoid the J x B drift of the plasma which introduces non-uniformity at the extraction plane. This proved a severe problem in the dipole version[7] and manifests at fields in excess of about 0.5mT. In the supercusp geometry there is also a filter at the ends of the source so that the drift continues, creating a "racetrack" situation where the J x B drift forms a closed ring and the plasma remains moderately uniform [8]. The distance between the filter and the extraction grid can be varied as an experimental tool.

It is important to note that for the purposes of comparing the model with experiment in which significant beam current is extracted, the source gas pressure input to the model must be adjusted to account for the gas streaming from the source in the form of beam ions. In the calculations that follow this is particularly relevant to the JET PINIs for which beam currents up to 60A are extracted during generation of the species data. Only gas flow into the source is measured, so the gas pressure was calculated by subtracting the beam equivalent flow from the measured value and then calculating the gas conductance out of the source assuming molecular flow. At 60A, the beam equivalent flow is approximately 50% of the gas flow.

*9.1     The "10×10" source*

This is a cubical magnetic multipole source (of 10 inch dimensions, hence the name) with a filter formed by small bar magnets inside water cooled tubes. By varying the magnets, Pincosy [7] created several magnetic filter geometries. The ion species were measured by a magnetic momentum analyser and the plasma properties were derived from Langmuir probes. Figure 3 shows the source species for a deuterium discharge as a function of ion current density at the extraction plane without a magnetic filter and for comparison the output from the model of this source.

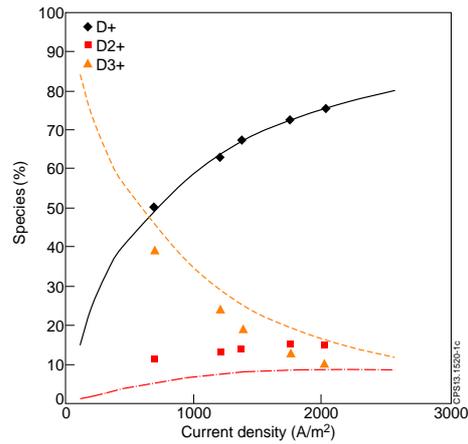

**Figure 3** The ion species measured in deuterium in the LBL "10 x 10" source and the output of the model for the case with no magnetic filter. The lines are the model calculations.

The addition of a magnetic filter to give average fields of 1.7 mT and 3.5 mT increases the deuteron fraction as shown in figures 4 and 5, again with the results of the model for comparison. Figures 3 to 5 indicate qualitative agreement between experiment and the model, which is able to reproduce the species variations observed in the discharge. There is a tendency to over-estimate the improvement in the deuteron yield with increasing filter field but the main trend of the improvement in deuteron fraction is clear. The deuteron fraction increases from zero discharge current density and the rate of rise and its limiting value at high arc currents follow the experimental data closely.

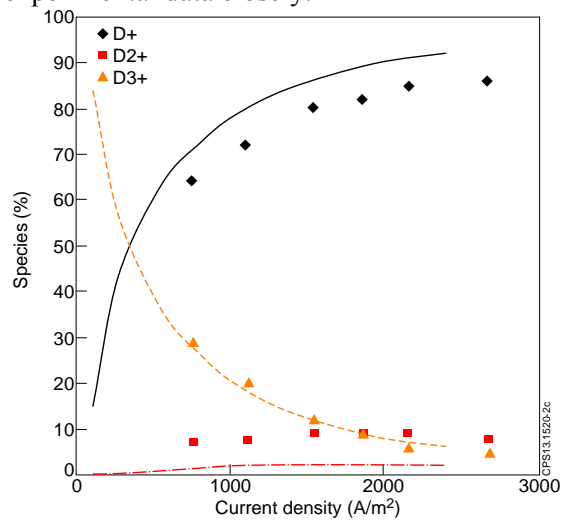

**Figure 4** The ion species measured in deuterium in the LBL "10 x 10" source and the output of the model for the case with dipole magnetic filter of average value 1.7mT. The lines are the model calculations.

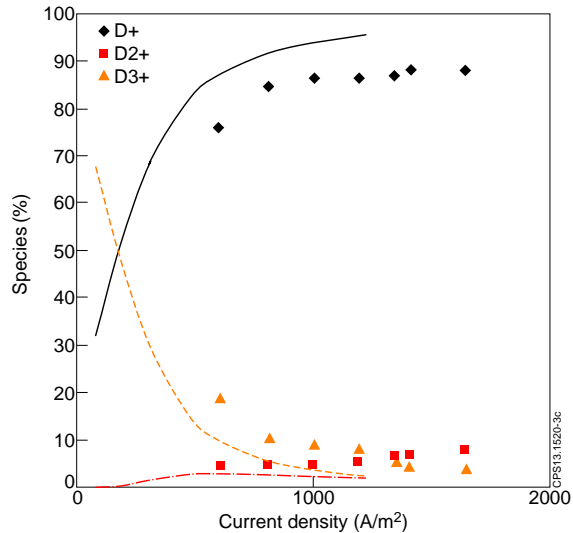

**Figure 5** The ion species measured in deuterium in the LBL "10 x 10" source and the output of the model for the case with dipole magnetic filter of average value 3.5mT. The lines are the model calculations.

The filter field increases the plasma density in the driver region where the filaments are situated and inhibits the plasma flow to the extraction plane, hence the truncation of the data at lower current density in figure 5. The primary electrons cross the filter field by collisions, losing energy in the process, which serves to reduce the plasma temperature. These effects are illustrated in figure 6 for a 3.5mT filter field in a 35kW deuterium discharge. The data are derived from Langmuir probe measurements. The electron temperature at the extraction plane is still quite high in this source and rises across the filter plane. However the model cannot follow the rise observed experimentally and gives a result 1eV less than experimentally observed in the driver region.

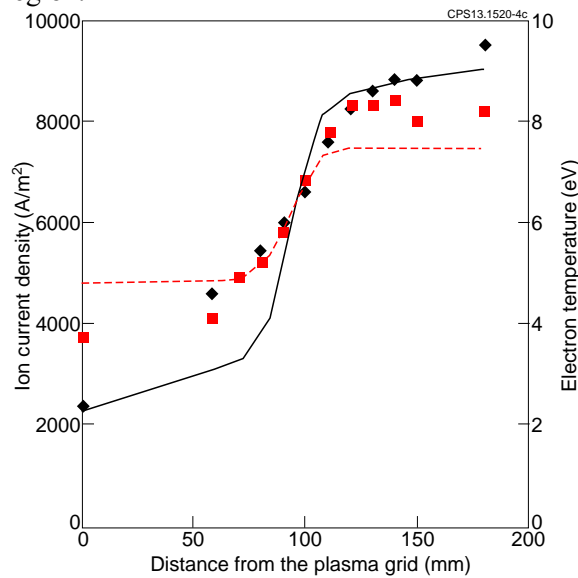

**Figure 6** The ion current density, $j_+$ ♦ and electron temperature, $T_e$, ■ as functions of distance from the plasma grid (origin) in deuterium in the LBL "10 x 10" source with dipole magnetic filter of average value 3.5mT. The data points are derived from Langmuir probe measurement. The solid line is current density and the broken curve electron temperature from the model calculations.

It is shown in the second paper that the main contributing factor to deuteron fraction is the large rise in current density in the driver region (about 4 fold) which increases molecular gas dissociation by electron collisions and hence direct deuteron production by atomic ionization. Molecular ion dissociation (which was thought to be the main reason for the increase [7]) is still present but is a lesser contributor

*9.2   The JET PINI source*

The PINI source used on the JET tokamak as a neutral beam injector for heating the toroidal plasma has interior dimensions 0.55x0.31x0.21m. The usual discharge gas is deuterium and there are extensive measurements of the beam fractions using a Doppler shift diagnostic which measures the intensity of full energy, half energy and one third energy $D^0$ atom velocities. The latter two particles originate from break-up of the $D_2^+$ and $D_3^+$ ions after ion-gas molecule collisions in the neutraliser of the injector. The intensities of these two $D^0$ fractions can be traced back to the original ion species in the discharge [2].

Without a filter field the maximum deuteron fraction approaches 75% with reasonable quantitative agreement of the model as can be seen in figure 7. With hydrogen, see paper II, the proton fraction is lower, reaching a maximum value of only 60%.

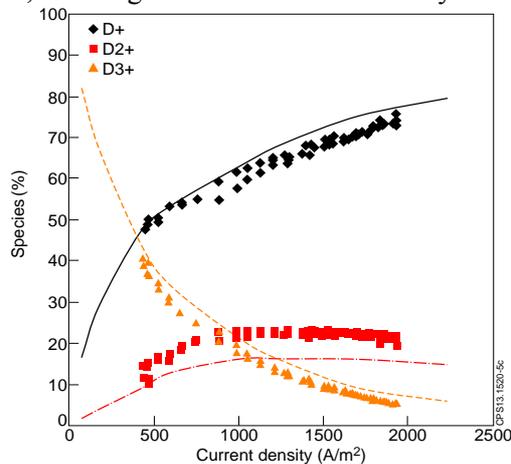

**Figure 7** Species ratios for the JET PINI source operated in deuterium without a filter field. The lines are the model calculations.

The supercusp is formed by arranging a net north pole on the backplane of the source (behind the filaments) and a ring of net south poles near the extraction plane so that the field has a form like a "circus tent" with the filaments on the outside of the "tent" (see figure 2). This minimises the uniformity problems encountered by dipole filter used by Pincosy, et al [7] arising from J × B drift orthogonal to the filter field and the source central axis. It was a successful approach [5, 6] as the non-uniformity was less than 5% of the plasma current density over the entire aperture array of 450×200 mm². Despite the filter field being at an angle of about $40^0$ to the source axis, the model works well in this situation, providing the field is reduced by the cosine of this angle, as is shown in figure 8. In this instance, the deuteron fraction reaches 91%.

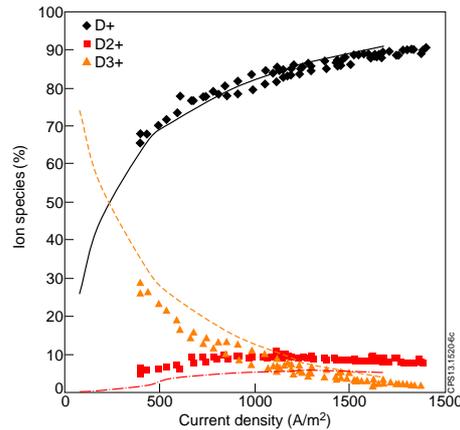

**Figure 8** Species ratio for the JET PINI source operated in deuterium with a supercusp magnetic filter field configuration. The lines are the model calculations.

Another test of the model that can be applied to the PINI source is calculation of the discharge efficiency; i.e. the current density available at the extraction plane for a given arc current as measured at the arc power supply (this is larger than the emission current used in the model due to the back ion bombardment of the cathode). This is shown in figure 9 below for a deuterium discharge with no filter field and with a supercusp filter:

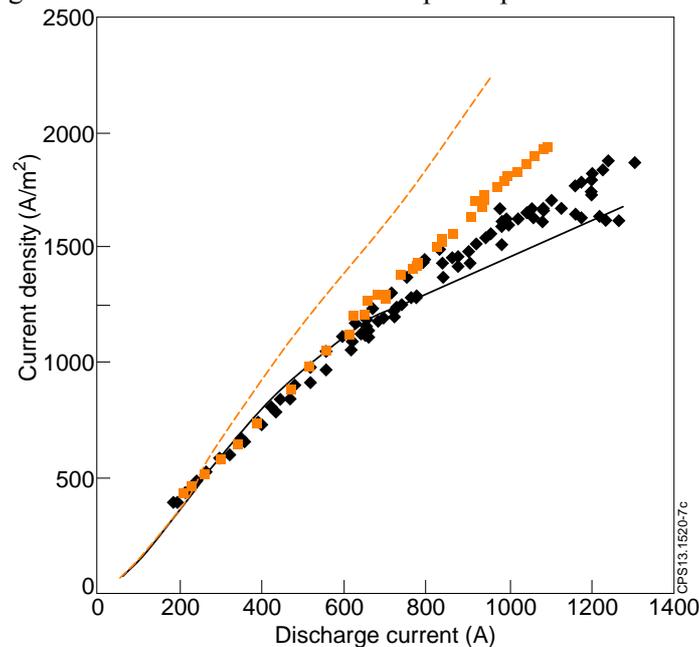

**Figure 9** The current density at the extraction plane as a function of discharge current, an indication of source efficiency. All data is for deuterium discharges: experimental data ■ source with no filter, ♦ source with filter, modelling --- no filter, ─ with filter.

The efficiency is noticeably lower for the supercusp source than the filterless source, a result that is also noted for the 10 × 10 source [7]. The main cause of the lower efficiency is the result of the lower plasma transport across the magnetic filter. As almost all of the ionization by fast primary electrons occurs in the driver region, near the filaments, the extraction plane becomes isolated from the ionization region and the plasma density and current density falls. The driver region, in contrast, shows a rapid rise in plasma density as the loss areas in the driver region (the filament structure and anode cusp lines) are small relative to the plasma grid area. This is seen in figure 6 for the 10 × 10 source, where the driver current density is much higher than at the extraction plane. The cusp losses, which are normally trivial, become significant at higher filter fields. The arc efficiency of very strong filters is even lower and will be discussed in paper two

*9.4  Variants of the JET PINI source*

Several variants of the Jet PINI source have been developed including a version with a much greater depth which tests the effects of increasing the distance over which the molecular ions can be broken up both with and without a filter field and a version used to create negative ions [8] which has a much stronger supercusp filter (see paper II).

The long PINI, i.e. the source of greater depth, which is 270 mm deep compared with the standard design of 200mm depth, has the same filter field in the same position relative to the backplane of the source, so that the filter field is now further from the extraction plane. This does increase the deuteron yield as shown in Fig 10:

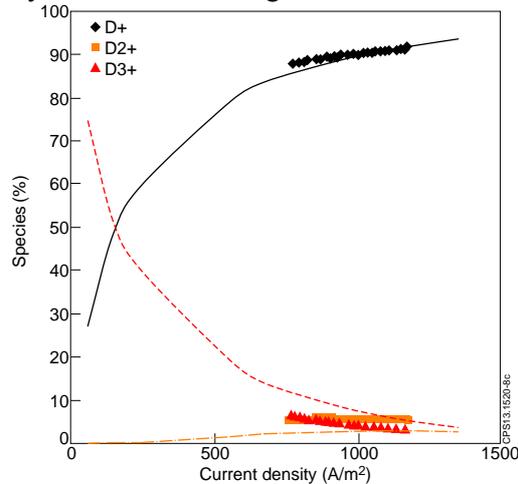

**Figure 10** Species ratios for the long JET PINI source operated in deuterium with a supercusp filter field. The lines are the model calculations.

However the increase in deuteron yield from 90% to 93% is at the expense of a decrease in discharge efficiency from 1.55A/m$^2$/A of discharge current to about 0.95 A/m$^2$/A of discharge current by comparison with figure 8. For this reason, the long PINI was not a success and a similar result was reported by Pincosy [7].

**10. Discussion and Conclusion**

In conclusion a model of the operation of magnetic multipole sources has been presented that provides reasonably good agreement with experimental values of the species fractions of positive deuterium ions and the discharge efficiency and is also compatible with the production of H$^-$ (D$^-$) ions by volume production. This model is based on classical plasma transport through a magnetic field and includes all of the relevant collision cross-sections, many of which have strong temperature dependence. The only input data is the source geometry, discharge voltage and current and the gas pressure. The model can be used to explore the mechanisms controlling the ion species yields in all the hydrogenic isotopes, this will be the subject of the second paper and to predict tritium yields for subsequent tritium injection experiments on JET, in a third paper. The introduction of surface production of negative ions at the plasma grid will be the subject of a subsequent paper. The model is, in principle, extendable to other gases by changing the relevant cross-sections and reaction rates, for example molecular gases such as boron trifluoride where the molecular ions are also broken up by low energy electron impact.

**Acknowledgement**
This work was funded by the RCUK Energy Programme under grant EP/I501045 and the European Communities under the contract of Association between EURATOM and CCFE.